\documentclass[proof]{pasj01}
\draft 
\Received{$\langle$reception date$\rangle$}
\Accepted{$\langle$acception date$\rangle$}
\Published{$\langle$publication date$\rangle$}
\begin{document}

\title{AN INTERMEDIATE VELOCITY HI CLOUD FALLING TO THE GALACTIC DISK; POSSIBLE EVIDENCE FOR LOW METALLICITY HI GAS ORIGINATED OUTSIDE THE GALACTIC DISK}
\author{Yasuo \textsc{Fukui}\altaffilmark{1}, Masako \textsc{Koga}\altaffilmark{1}, Shohei \textsc{Maruyama}\altaffilmark{1}, Takahiro \textsc{Hayakawa}\altaffilmark{1}, Ryuji \textsc{Okamoto}\altaffilmark{1}, Hiroaki \textsc{Yamamoto}\altaffilmark{1}, Kengo\textsc{Tachihara}\altaffilmark{1}, Robin \textsc{Shelton}\altaffilmark{2} and Manami \textsc{Sasaki}\altaffilmark{3}}%
\altaffiltext{1}{Department of Physics, Nagoya University, Chikusa-ku, Nagoya 464-8602, Japan}
\altaffiltext{2}{Department of Physics and Astronomy, University of Georgia, Athens, GA 30602, USA}
\altaffiltext{3}{Dr. Karl Remeis-Sternwarte, Erlangen Centre for Astroparticle Physics, Friedrich-Alexander-Universit\"at Erlangen-N\"urnberg, Sternwartstra{\ss}e 7, D-96049 Bamberg, Germany}
\email{fukui@a.phys.nagoya-u.ac.jp}


\maketitle

\begin{abstract}
We found that an intermediate velocity cloud (IVC) IVC 86$-$36 in H{\sc i} 21 cm emission shows a head-tail distribution toward the Galactic plane with marked parallel filamentary streamers, which is extended over 40 degrees in the sky.
The distance of IVC 86$-$36 is constrained to be less than $\sim 3$ kpc from absorption of a background star as determined from optical spectroscopy.
There is a bridge feature in velocity between the IVC and the local ISM with velocity separation of $\sim 50$ km s$^{-1}$, which may indicate dynamical interaction of the IVC with the disk.
If the interaction is correct, the distance estimate $d$ of the IVC ranges from 200 pc to 3 kpc, and the mass of the IVC head is estimated to be $7\times 10^3(d/1\mathrm{kpc})^2\, M_{\solar}$.
The IVC shares similar properties to the Smith cloud located at 12 kpc, including the head-tail distribution, streamers, and bridge feature, while the mass of the IVC is less than $\sim 0.1$ of the Smith cloud.
A comparison between the H{\sc i} and the \textit{Planck}/\textit{IRAS} dust emission indicates that the dust emission of IVC 86$-$36 is not detectable in spite of its H{\sc i} column density of $2\times 10^{20}$ cm$^{-2}$, indicating low metalicity of IVC 86$-$36 by a factor of $\ltsim 0.2$ as compared with the solar neighbor.
We conclude that IVC 86$-$38 is an infalling cloud which likely originated in the low-metallicity environment of the Galactic halo or the Magellanic system. 
\end{abstract}

%% INTRODUCTION
\section{Introduction}\label{sec:introduction}

The Galactic halo harbors a number of high- to intermediate-velocity H{\sc i} clouds (HVCs and IVCs) as revealed by 21 cm H{\sc i} observations \citep{1997ARA&A..35..217W, 2004ASSL..312.....V}.
It has been discussed that they may be external clouds falling onto the Galaxy and are the remnant of the building blocks of the Galaxy, expelled clouds of Galactic fountains, and H{\sc i} gas stripped from the Magellanic Cloud or the other objects in the Local Group \citep{1997ARA&A..35..217W, 1999ApJ...514..818B, 1999A&A...341..437B, 2002ApJS..140..331L, 2004MNRAS.355..694M, 2006ApJ...646L..53C}.
 Some of the clouds have cometary morphologies, suggesting that they are interacting with Galactic halo material \citep{1990ApJ...356..130M, 2000A&A...357..120B, 2004ASSL..312..251B, 2007ApJ...656..907P}.
It is important to better understand the physical and kinematical states of the HVCs and IVCs in order to elucidate the history and physics of the Galactic disk-halo interaction. 

The Smith Cloud \citep{1963BAN....17..203S}, located toward ($l$, $b$) = ($\timeform{38.67D}$, $\timeform{-13.41D}$) is a unique object among the HVCs \citep{1984ApJ...283...90L, 1997ARA&A..35..217W}.
It was shown that the Smith cloud is interacting with the disk H{\sc i} gas based on 21 cm H{\sc i} observations \citep{2007AAS...21114801L}.
The cloud shows unique signatures of interaction with the Galactic halo material, including filamentary parallel features with a head-tail structure as revealed by the Green Bank Telescope (GBT).
Multiple origins of the cloud have been proposed in the literature \citep{1998MNRAS.299..611B, 2004PASJ...56..633S}.
A recent scenario under intensive attention is a cloud falling through the Galactic halo after a passage through the disk $\sim$ 70 Myr ago, where dark matter may play a role in confining the cloud \citep{2009ApJ...707.1642N}, whereas the hypothesis is not supported by the hydrodynamical numerical simulations \citep{2016ApJ...816L..18G}.
The Smith cloud is located below the galactic plane ($z\sim 3$ kpc) at a distance of $12.4\pm 1.3$ kpc from the sun and the total H{\sc i} mass is estimated to be more than 10$^6$ $M_{\solar}$.
The metallicity of the Smith cloud is measured to be sub-solar but is still not inconsistent with the metallicity of the Galactic disk \citep{2016ApJ...816L..11F}.
Its origin may be explained by the Galactic fountain model where the disk H{\sc i} gas is ejected out of the plane by supershells, the collective effects of supernovae.

IVCs are already well studied in H{\sc i} at low resolution.
The Pegasus-Pisces Arch \citep{2001ApJS..136..463W} harbors several IVCs with elongated distributions.
In the course of a study of the region \citep{2014ApJ...796...59F}, we made a detailed analysis of the H{\sc i} data taken as part of the GALFA survey \citep{2011ApJS..194...20P} with the Arecibo telescope at $\timeform{4'}$ resolution.
In this paper we present the results of the H{\sc i} analysis toward an IVC in the Pegasus-Pisces Arch.

%% Results
\section{Results of the H{\sc i} distribution}
\subsection{Observational data}\label{subsec:obsdata}

We use the archival datasets of the Galactic Arecibo L-Band Feed Array (GALFA)-H{\sc i} survey.
The GALFA-H{\sc i} survey data \citep{2011ApJS..194...20P} is taken with the Arecibo 305 m radio telescope. Its angular resolution is $\timeform{4'}$ and rms noise level is 60--140 mK at 0.2 km s$^{-1}$ velocity resolution in $T_{\rm MB}$ scale which is the Main Beam antenna temperature.

\subsection{Spatial distribution of the IVC}

\begin{figure}
 \begin{center}
 \includegraphics[scale=1.0]{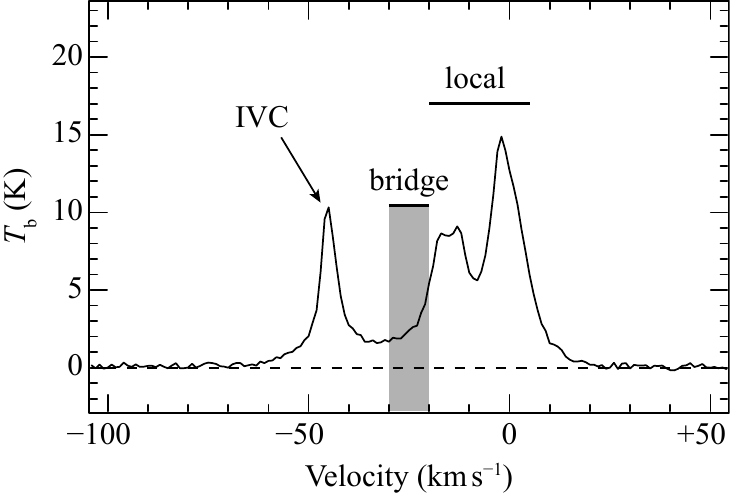}
 \end{center}
 \caption{A typical profile of the H{\sc i} in the MBM53--55/IVC 86$-$36 region. The component of the IVC, bridge and local gas are represented in the figure.}\label{fig:typicalprofile}
\end{figure}

We present the intermediate velocity H{\sc i} cloud at ($l$, $b$) = ($\timeform{86D}$, $\timeform{-36D}$), IVC 86$-$36, in the Pegasus-Pisces Arch, where the name in Galactic coordinates is tentatively adopted for convenience in the present paper.
The velocity with respect to the local standard of rest (LSR) is given in the present paper.
Figure \ref{fig:typicalprofile} shows a typical H{\sc i} profile toward the IVC in velocity and shows that there are two velocity components; one is the local emission in a low velocity range from $-$20 km s$^{-1}$ to +10 km s$^{-1}$ and the other is peaked at large negative velocity around $-$45 km s$^{-1}$.
This negative velocity component is IVC 86$-$36, which is apparently linked with the local H{\sc i} gas by a smooth bridge-like feature as described later.

\begin{figure*}
 \begin{center}
  \includegraphics[scale=1.0]{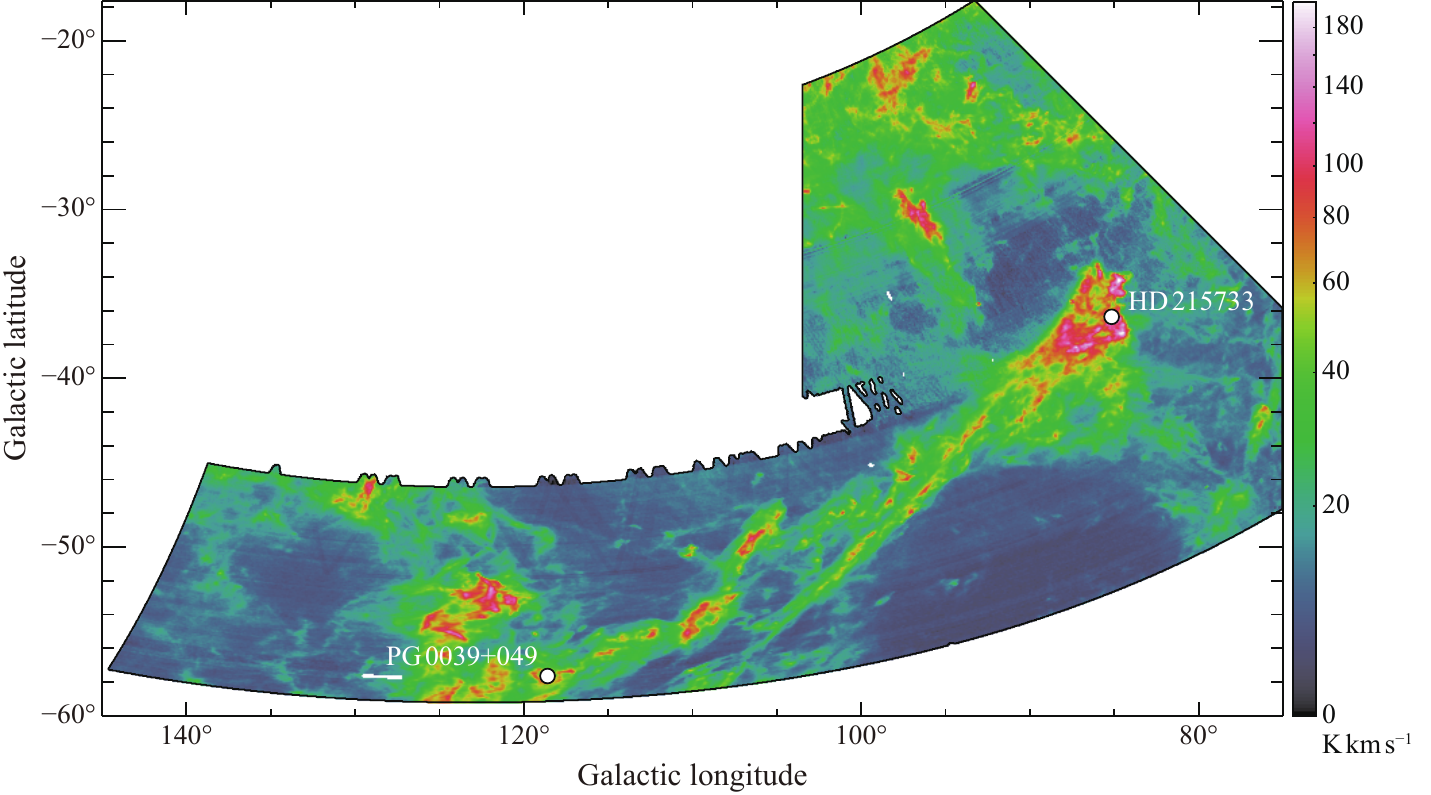}
 \end{center}
 \caption{The H{\sc i} distribution of the IVC 86$-$36 integrated over a velocity range from $-$80 km s$^{-1}$ to $-$30 km s$^{-1}$. The position of PG0039+049 and HD215733 is shown in the figure.}\label{fig:largescaleHI}
\end{figure*}

\begin{figure*}
 \begin{center}
  \includegraphics[scale=1.0]{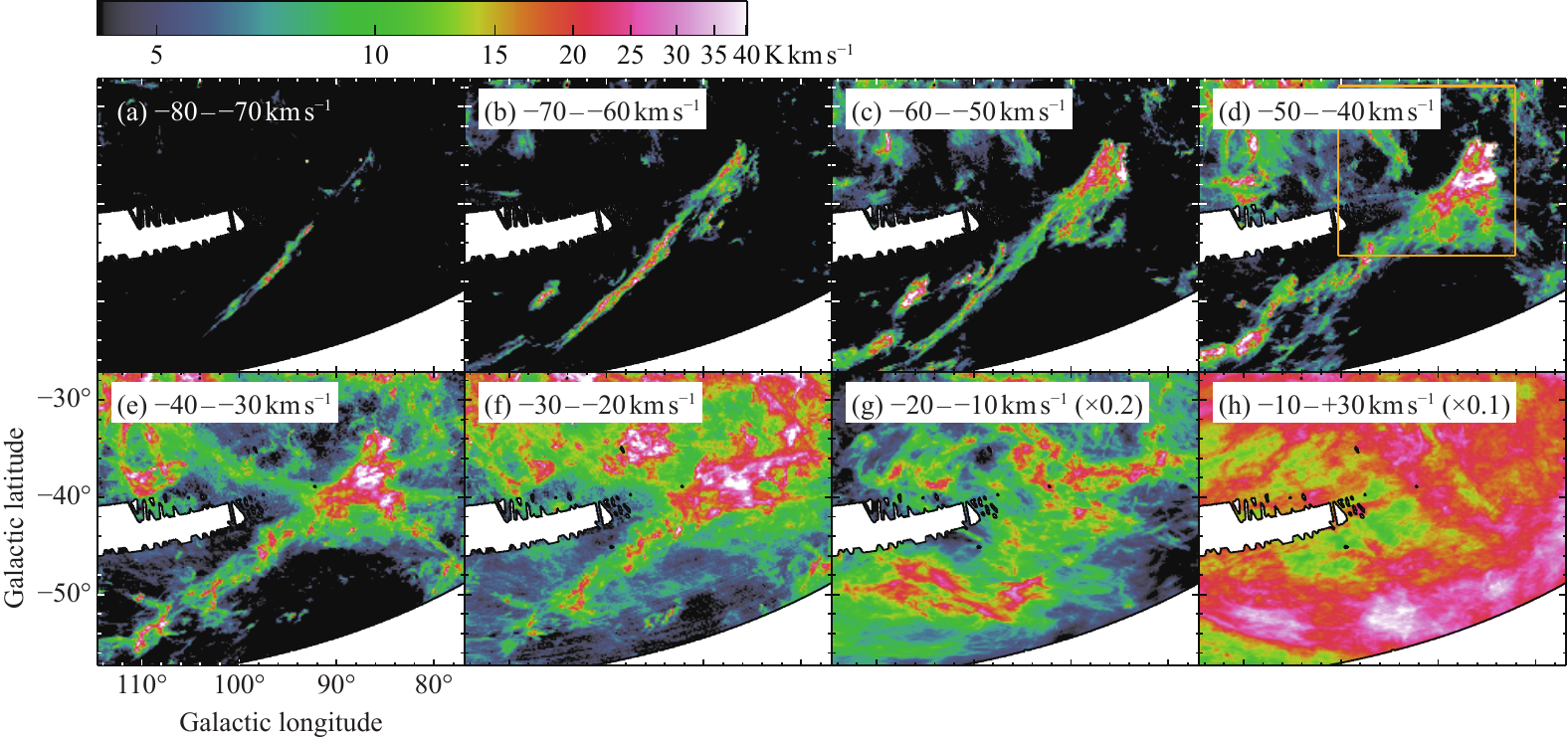}
 \end{center}
 \caption{The velocity channel maps of the H{\sc i} every 10 km s$^{-1}$ from $-$80 km s$^{-1}$ to +30 km s$^{-1}$. The velocity-integarted intensity in (g) and (h) is multiplied by 0.2 and 0.1, respectively. The bounding box in (d) indicates the region shown in Figure \ref{fig:dust_HI_maps}.}\label{fig:channelmap}
\end{figure*}

\begin{figure}
 \begin{center}
  \includegraphics[scale=1.0]{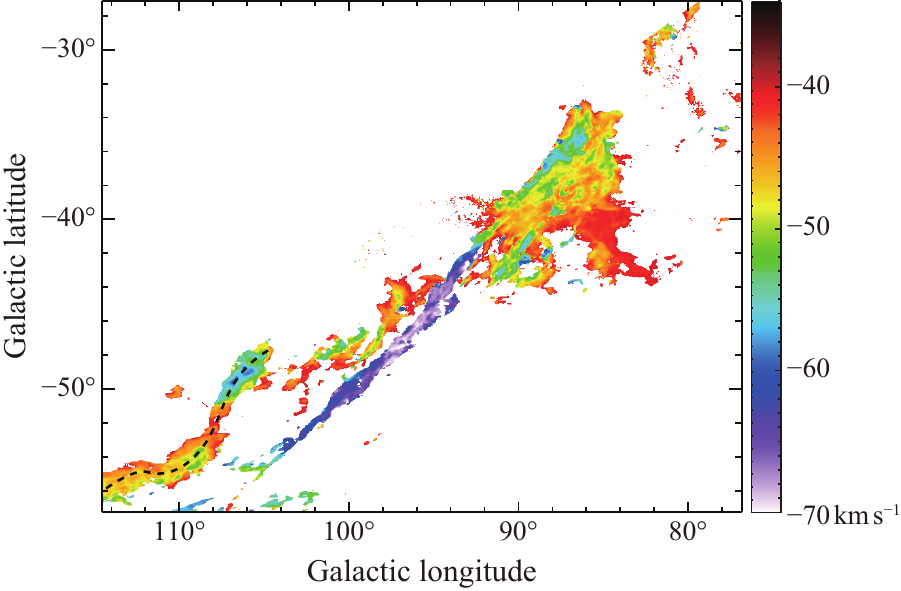}
 \end{center}
 \caption{The first moment map of the H{\sc i} calculated in a velocity range from $-$70 km s$^{-1}$ to $-$35 km s$^{-1}$. The winding structure described in the text is shown by a dashed line.}\label{fig:momentmap}
\end{figure}

Figure \ref{fig:largescaleHI} shows a large scale distribution of IVC 86$-$36, which is extended as a thin and straight feature in a region of ($l$, $b$) = ($\timeform{85D}$--$\timeform{120D}$, $\timeform{-60D}$--$\timeform{-30D}$).
The velocity of the IVC ranges from less than $-$60 km s$^{-1}$ to $-$20 km s$^{-1}$.
Figure \ref{fig:channelmap} shows an image of the bright H{\sc i} 21 cm emission from $-$80 km s$^{-1}$ to +30 km s$^{-1}$ as velocity channel distributions.
IVC 86$-$36 is seen from $-$80 km s$^{-1}$ to $-$30 km s$^{-1}$ as the straight feature from ($l$, $b$)=($\timeform{84D}$, $\timeform{-34D}$) to ($\timeform{110D}$, $\timeform{-55D}$).
The angle of the elongated IVC relative to the Galactic plane is $\sim$45 degrees.
IVC 86$-$36 has a head-tail distribution whose head, the compact brightest part, lies toward ($l$, $b$) = ($\timeform{84D}$--$\timeform{90D}$, $\timeform{-40D}$--$\timeform{-34D}$).
The bridge is seen in $-30$ km s$^{-1}$ -- $-20$ km s$^{-1}$.
At velocity from $-$20 km s$^{-1}$ to 10 km s$^{-1}$ the local H{\sc i} gas is dominant and we do not see the IVC.
Figure \ref{fig:momentmap} presents the first moment distribution calculated in a velocity range from $-$70 km s$^{-1}$ to $-$30 km s$^{-1}$, showing that the red-shifted tail at velocity larger than $-$60 km s$^{-1}$ has ``winding'' distribution as depicted by a dashed line in Figure \ref{fig:momentmap}, whereas the blue-shifted tail at velocity smaller than $-$60 km s$^{-1}$ holds the straight distribution.

The present region includes high latitude molecular clouds MBM53--55 \citep{1985ApJ...295..402M}.
H{\sc i} distribution was already studied at lower resolution of $\timeform{30'}$ by \citet{2003ApJ...592..217Y} for a velocity range of $-20$ km s$^{-1}$ to $+10$ km s$^{-1}$, and the major velocity component of H{\sc i} associated with the CO cloud was revealed, which correspond to the H{\sc i} from $-11$ km s$^{-1}$ to $+1$ km s$^{-1}$.

\subsection{A possible kinematic signature of interaction between the IVC and the local gas}

\begin{figure*}
 \begin{center}
  \includegraphics[scale=1.0]{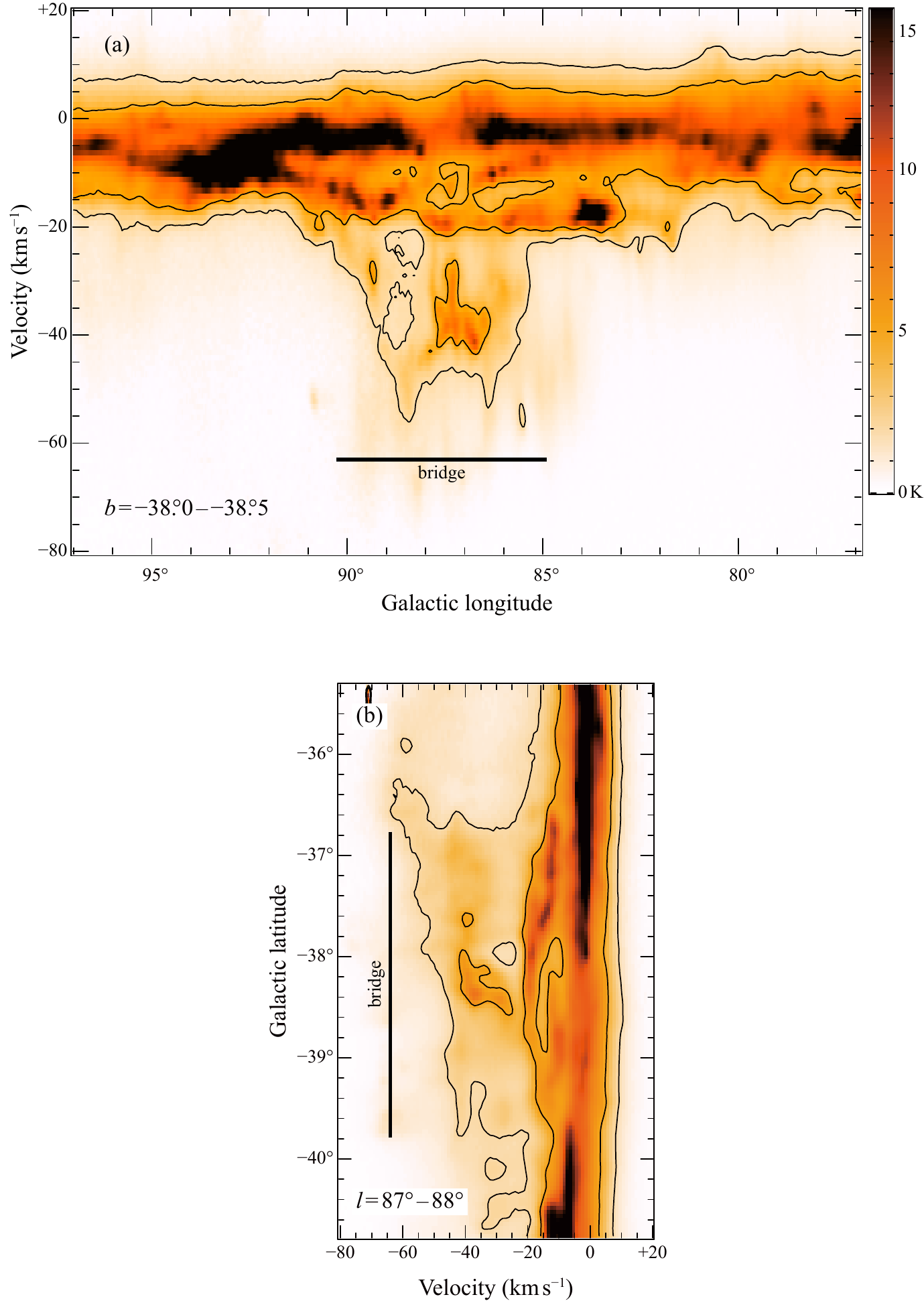}
 \end{center}
\caption{The Galactic longitude-velocity diagram (a) and the velocity-Galactic latitude diagram (b) in the H{\sc i}. The intensity is averaged over each Galactic latitude and longitude in (a) and (b), respectively. The contour levels in both (a) and (b) are 2 K and 5 K. The position of bridge feature is shown in the figures.}\label{fig:pvdiagrams}
\end{figure*}

In order to obtain an insight into the relationship between the IVC and the disk, in Figure \ref{fig:pvdiagrams} we present position-velocity diagrams toward the IVC head.
In Figure \ref{fig:pvdiagrams}(a) we present a Galactic longitude-velocity diagram toward the IVC head, where the bridge feature is found at $-$30 km s$^{-1}$ -- $-$20 km s$^{-1}$ for a Galactic longitude range from $\timeform{85D}$ to $\timeform{90D}$ over the extent of the intense part of the IVC head at $\sim -40$ km s$^{-1}$.
In Figure \ref{fig:pvdiagrams}(b) a Galactic latitude-velocity diagram in the IVC head is shown.
We find the bridge feature in a Galactic longitude range from $\timeform{-40D}$ to $\timeform{-37D}$.
Numerical simulations of two colliding clouds indicate formation of such bridges between the two clouds \citep{2014ApJ...792...63T, 2018ApJ...859..166F}, and the distribution of the bridge is consistent with that IVC 86$-$36 is interacting with the disk H{\sc i}.
It is possible that the H{\sc i} disk have a vertical extent of more than 3 kpc as shown by the bridge features toward the Smith cloud \citep{2007AAS...21114801L}.
The similar extents of the bridge and the IVC head suggest physical connection of the IVC and the local gas.

\section{Distance and mass of the IVC}\label{sec:d_and_m_of_IVC}
\subsection{Distance}
It is difficult to derive the distance of IVCs which do not follow the Galactic rotation and probably have a significant radial motion to the Galactic disk.
In the following we present and discuss a few possible ways of distance estimate and their uncertainties. 

\textbf{Stellar absorption:} IVCs are generally believed to be closer to the Galactic plane than HVCs and are located at a $z$-distance around 1 kpc (e.g., \cite{2001ASPC..240..529C, 2017ASSL..430...15R}).
A secure limit for the distance of IVC 86$-$36 is obtained by absorption toward background stars.
HD 215733 is located toward the head of IVC 86$-$36 at Galactic coordinates of ($l$, $b$)=($\timeform{85.2D}$, $\timeform{36.4D}$).
The distance of HD 215733 is estimated to be 2900 pc based on a spectral type of B1 II, $V = 7.3$, $E(B-V) = 0.10$, and $M_V = -5.3$ \citep{1976ApJ...205..419W, 1987A&AS...71..119M, 1970A&A.....4..234F, 1968ApJS...17..371L}.
Recent measurements by GAIA obtained $3.5\pm 0.9$ kpc as a distance of HD215733, which is consistent with the previous determination above.
Another constraint on the distance was obtained to be less than 1050 pc (see \cite{2001ApJS..136..463W}) by the optical absorption measurement toward a star PG0039+049 at ($l$, $b$) = ($\timeform{119D}$, $\timeform{58D}$) in part of the tail of IVC 86$-$36 \citep{2001ApJS..136..463W, 1994A&A...292..261C}.

\textbf{\boldmath{$A_{V}$} estimation by stars observed with GAIA:} H{\sc i} column density $N_\mathrm{HI}$ is estimated to be $2\times 10^{20}$ cm$^{-2}$ for the H{\sc i} peaks of IVC 86$-$36 by assuming that the H{\sc i} is optically thin.
This $N_\mathrm{HI}$ corresponds to $A_\mathrm{V}=0.1$ mag \citep{2011piim.book.....D} if the Galactic gas to dust ratio is assumed to be the same in IVC 86$-$36.
GAIA DR2 presented the most extensive catalog of stars with their parallax and other stellar properties \citep{2018A&A...616A...1G}.
We made an analysis of $A_{V}$ toward IVC 86$-$36 as described in the Appendix and found that $A_{V}$ is enhanced by $\sim 2$ mag at a distance of 100--200 pc, which corresponds to the local molecular clouds MBM53--55 with an H{\sc i} envelope.
This is consistent with the results by PS1 photometry \citep{2014ApJ...786...29S}.
For the rest of the line of sight we found no enhanced $A_{V}$ with a upper limit of $A_{V}\sim 0.3$ mag within $\sim 1$ kpc.
The GAIA DR2 data are not sensitive enough to detect such small $A_{V}$ of the IVC and the distance estimate for IVC 86$-$36 is not constrained with GAIA DR2. 

\textbf{X-ray shadow:} Another possibility is to use the soft X-rays at around 1/4 keV observed with ROSAT to identify shadow due to X-ray absorption.
Such X-ray absorption by the ISM is mainly due to hydrogen and helium with a possible contribution of heavy atoms at high energy above 0.5 keV.
This method was successful in identifying the X-ray shadow of the Draco cloud \citep{1991Natur.351..629B, 1991Sci...252.1529S}.
We find possible shadow candidates toward MBM53--55 in the catalog of \citet{2000ApJS..128..171S} as given in Appendix 2.
Some of the cataloged shadows may correspond to IVC 86$-$36, whereas $W_\mathrm{HI}$ of the IVC is a factor of 3--5 smaller than that of the local H{\sc i} gas.
It is possible that the soft X-ray background is mainly due to the Local Bubble distributed within 200 -- 300 pc, making the shadow not very distinct.
The 3D distribution of the soft X-ray emitting hot gas of the Local Bubble is not accurately established.
A quantitative analysis of the shadows may include large uncertainties at best and we did not attempt to derive a distance of IVC 86$-$36 by using the X-ray shadows.

\textbf{Morphology of IVC 86$-$36:} The IVC morphology may provide another hint on the location of the cloud.
The negative velocity of IVC 86$-$36 indicates that the IVC is approaching the sun.
If the tail was created by the interaction between the head and the halo material, it is likely that the tail is placed at a larger distance than the head.
The head then conflicts with 1 kpc of PG0039+049, if the intermediate velocity gas toward PG0039+049 is a physical extension of IVC 86$-$36.
The distance 3 kpc may not be favored in the configuration.

In summary, the distance of IVC 86$-$36 is still uncertain observationally and we shall consider three distances, 200 pc, 1 kpc, and 3 kpc, in the following discussion. 

\subsection{Cloud size and mass}
We assume in the following that the tail of the cloud is at the same projected distance with the head in the disk for simplicity.
Under the assumption, the head is at a distance of $d$ and the edge of the tail is at $d \cos(\timeform{36D})/\cos(\timeform{60D})=1.62d$.
By considering the extremely straight distribution of IVC 86$-$36 in the sky, we assume that the cloud is straight in the 3-dimensional space.
This is a natural configuration if the tail was produced by the interaction between the IVC head and the halo material at a relative velocity of $\sim 100$ km s$^{-1}$.
The real length of the cloud $L$, separation between IVC head and tail, is calculated as $L=0.66 d$, where $d$ stands for the distance of the head.
The length of the tail is calculated as $L = 130$ pc, 660 pc and 2 kpc, for a distance of 200 pc, 1 kpc and 3 kpc, respectively.

The low brightness of the H{\sc i} emission suggests that the emission is optically thin \citep{2014ApJ...796...59F}, and we estimate the H{\sc i} mass under the optically thin approximation by equation (2).
\begin{equation}
M_\mathrm{HI}=\mu m_\mathrm{H}\sum\left[d^2\Omega X_\mathrm{HI} W_\mathrm{HI}\right]
\end{equation}
where $d$ is the distance of the IVC, $\omega$ is the solid angle per pixel, $X_\mathrm{HI}=1.823\times 10^{18}$ [cm$^{-2}$ K$^{-1}$ km$^{-1}$] s is a theoretical conversion factor between the integrated intensity ($W_\mathrm{HI}$) and the column density of H{\sc i}.
The total H{\sc i} mass is estimated to be $7300\times (d/1\,\mathrm{kpc})^2\, M_{\solar}$  including He mass, 1/3 of H (taken into account in = 1:33).
The mass of IVC 86$-$36 is estimated to be 260 $M_{\solar}$ to $7\times 10^4$ $M_{\solar}$, which is significantly smaller than that of the Smith cloud whose mass is more than $10^{6}$ $M_{\solar}$ \citep{2007AAS...21114801L}.

\section{Gas to dust ratio in IVC 86$-$36}
In order to pursue the origin of an IVC, it is important to measure the metallicity.
IVCs may have lower metallicity if they originate outside the disk, e.g., in the Magellanic system or in the Galactic halo.
Metal abundance in the gas phase of IVC 86$-$36 was measured by optical atomic absorption lines toward HD215733 by \citet{1997ApJ...475..623F} and depletion of some heavy elements are measured in a velocity range of the IVC.
These measurements, however, may not directly indicate the metallicity of IVC 86$-$36 because of the uncertainty in assessing the atomic ionization states and the interstellar depletion due to adsorption onto dust surfaces.

\begin{figure*}
 \begin{center}
 \includegraphics[scale=1.0]{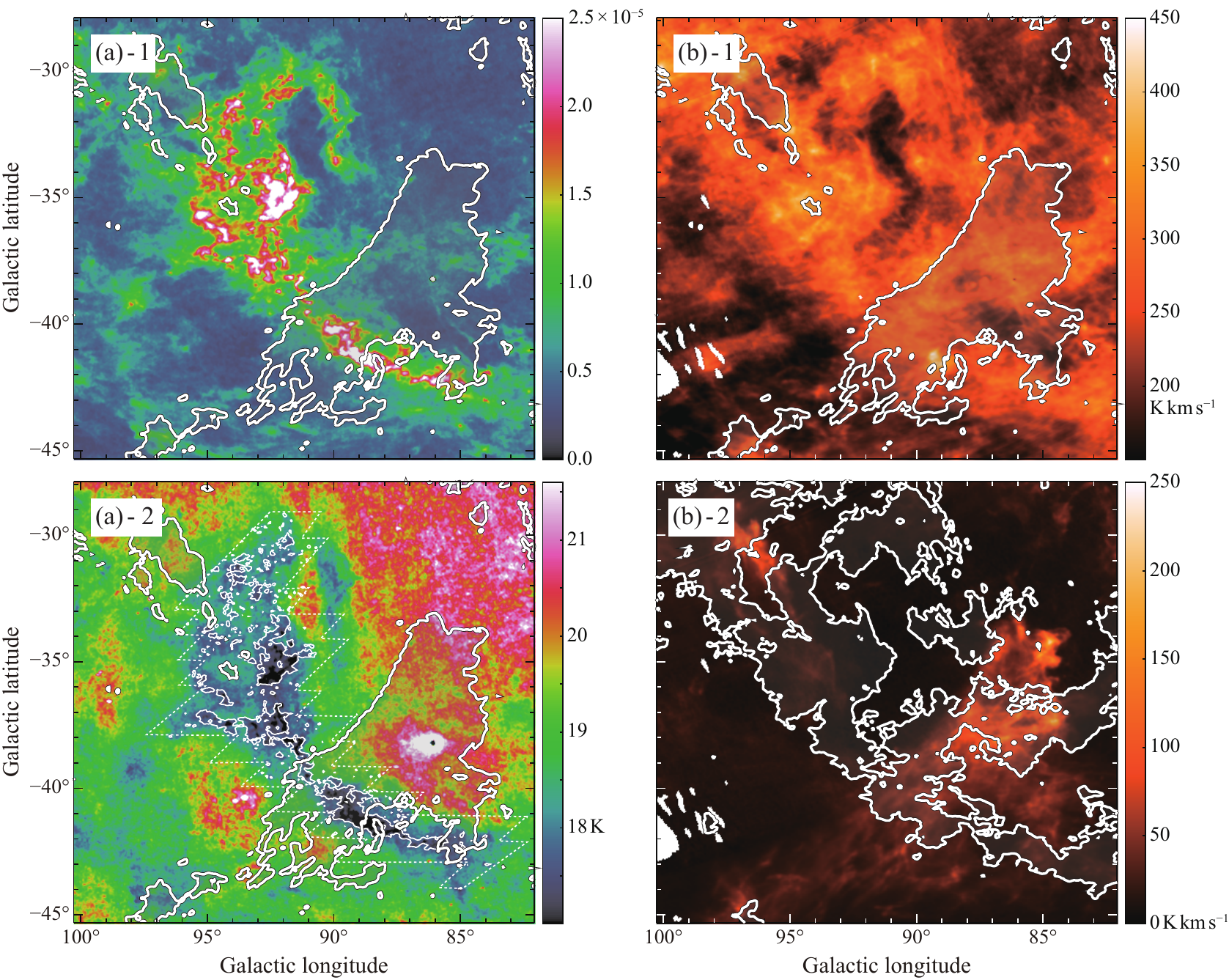}
 \end{center}
\caption{
(a) Spatial distributions of $\tau_{353}$ ((a)-1) and $T_\mathrm{d}$ ((a)-2) \citep{2014A&A...571A..11P}.
(b) Those of of $W_\mathrm{HI}$ integrated in a velocity range from $-30$ km s$^{-1}$ to $+30$ km s$^{-1}$ ((b)-1) and from $-60$ km s$^{-1}$ to $-30$ km s$^{-1}$ ((b)-2).
The gray-shades outlined by thick contours in panels (a)-1, (a)-2 and (b)-1 show the distribution of the IVC and that in panel (b)-2 show the local ISM.
The thin contours in panel (a)-2 show spatial distribution of $^{12}$CO (using the same dataset as Figure 1(a) of \cite{2014ApJ...796...59F}) integrated in a velocity range from $-12$ km s$^{-1}$ to +2 km s$^{-1}$.
}\label{fig:dust_HI_maps}
\end{figure*}

Another method to measure metallicity is the dust emission or extinction which reflects dust abundance.
The most sensitive observations of the dust emission at sub-mm wavelengths are obtained by the combined \textit{Planck}/\textit{IRAS} data.
Figure \ref{fig:dust_HI_maps} shows the distributions of radiation toward the region of IVC 86$-$36.
Figure \ref{fig:dust_HI_maps}(a)-1 shows the distribution of the dust optical depth at 353 GHz $\tau_{353}$ and Figure \ref{fig:dust_HI_maps}(a)-2 the distribution of the dust temperature $T_\mathrm{d}$.
These two quantities are derived by fitting a modified Planck function at three wavelengths of the \textit{Planck} data, 850, 550 and 350 {\micron} (corresponding to frequencies of 353, 545 and 857 GHz) \citep{2014A&A...571A...6P}, in addition to the \textit{IRAS} 100 {\micron} data \citep{1998ApJ...500..525S,2005ApJS..157..302M}.
The \textit{IRAS} data were smoothed to the beam size of the \textit{Planck} data \timeform{5'}.
The dust emissivity is expressed as $I_{\nu}=\tau_{353} B_{\nu}(T_\mathrm{d})(\nu/\mbox{353 GHz})^{\beta}$ where a coefficient $\beta$ were simultaneously fit with $\tau_{353}$ and $T_\mathrm{d}$.
For more details see \citet{2014A&A...571A..11P}.

Figure \ref{fig:dust_HI_maps}(b)-1 shows in gray scale the distribution of the integrated intensity of the local H{\sc i} emission in a velocity range of $-30$--$+30$ km s$^{-1}$, and Figure \ref{fig:dust_HI_maps}(b)-2 the distribution of the IVC H{\sc i} emission in a velocity range of $-60$--$-30$ km s$^{-1}$.
The white thick contours in Figures \ref{fig:dust_HI_maps}(a)-1, (a)-2 and (b)-1 show the outer contour of the IVC at an intensity level of 30 K km s$^{-1}$.
The white contours in Figure \ref{fig:dust_HI_maps}(b)-2 are the H{\sc i} intensity of the local ISM in Figure \ref{fig:dust_HI_maps}(b)-1 at an intensity level of 260 K km s$^{-1}$.
The thin contours in Figure \ref{fig:dust_HI_maps}(a)-2 shows the CO distribution at 3.4 K km s$^{-1}$ of the local ISM in a velocity range of $-12$--$+2$ km s$^{-1}$.
$\tau_{353}$ is a sum of the contributions of both the local ISM and the IVC.
The main H{\sc i} cloud (Figure \ref{fig:dust_HI_maps}(b)-1) that includes the high latitude molecular clouds MBM53--55 observed in CO (Figures \ref{fig:dust_HI_maps}(a)-2) is distributed in a velocity range of the local ISM from the northeast to the southwest and the distribution is similar to $\tau_{353}$ (Figure \ref{fig:dust_HI_maps}(a)-1).
The H{\sc i} distribution is extended outside the molecular cloud and shows weaker $\tau_{353}$ than the main cloud (Figures \ref{fig:dust_HI_maps}(a)-1 and \ref{fig:dust_HI_maps}(b)-1).
We note that IVC 86$-$36 (Figure \ref{fig:dust_HI_maps}(b)-2) shows no appreciable enhancement of $\tau_{353}$ (Figure 10(a)-1) or no correlated variation in $T_\mathrm{d}$ (Figure \ref{fig:dust_HI_maps}(a)-2) for $W_\mathrm{HI}$ of the IVC $\sim200$ K km s$^{-1}$.
$N_\mathrm{HI}$ of the IVC head corresponds to sub-mm dust optical depth $\tau_{353}=3\times 10^{-6}$ according to the works on the \textit{Planck}/\textit{IRAS} dust properties \citep{2014A&A...571A..11P, 2014ApJ...796...59F, 2015ApJ...798....6F}.
This suggests that IVC 86$-$36 has a significantly smaller dust optical depth than the local ISM. 

\begin{figure}
 \begin{center}
 \includegraphics[scale=1.0]{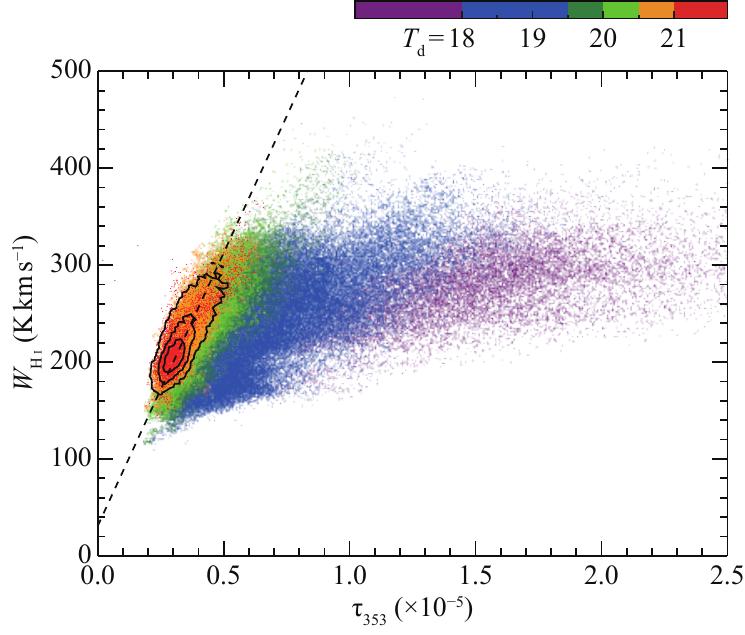}
 \end{center}
\caption{
Scatter plot between $\tau_{353}$ and $W_\mathrm{HI}$ for the local ISM ($W_\mathrm{HI}$ are integrated from -30 to +30 km s$^{-1}$).
Color represents $T_\mathrm{d}$ of each point.
The contours include 30\%, 60\%, and 90\% of data points with $T_\mathrm{d} > 20.5$K and the dashed line is linear regressions for the data points.
}\label{fig:353-WHI_localISM}
\end{figure}

Figure \ref{fig:353-WHI_localISM} shows a scatter plot between $\tau_{353}$ and $W_\mathrm{HI}$ of the local ISM for $\tau_{353}$ above $2\times 10^{-6}$, where the dust temperature derived from the \textit{Planck}/\textit{IRAS} data $T_\mathrm{d}$ is shown by the color code.
$T_\mathrm{d}$ is positively correlated with the H{\sc i} spin temperature $T_\mathrm{s}$ because the both temperatures reflect heating by UV of the ISRF, and $T_\mathrm{s}$ increases with the increase of $T_\mathrm{d}$; H{\sc i} heating by the photoelectric effect and dust heating by the UV absorption.
The both temperatures differ substantially from each other because collisional coupling between H{\sc i} and dust is too weak to equalize their temperatures in the usual CNM \citep{2011piim.book.....D}.
Due to the strong $T_\mathrm{d}$ dependence of the dust radiation energy as expressed by $T_\mathrm{d}^6$, $T_\mathrm{d}$ ranges from 16 to 22 K \citep{2014A&A...571A..11P} and $T_\mathrm{s}$ from 10 K to $10^{4}$ K (e.g., \cite{2018ApJ...860...33F}).
$\tau_{353}$ is a sum of the foreground ISM and the IVC.
In order to quantify the gas/dust ratio of IVC 86$-$36 we subtract the contribution of the foreground ISM and derive $\tau_{353}$ of the IVC.
In the present region \citet{2014ApJ...796...59F} showed that the H{\sc i} gas with $T_\mathrm{d}$ higher than 21 K is optically thin and $W_\mathrm{HI}$ is proportional to $N_\mathrm{HI}$ with a small dispersion of less than 10 \% in that $T_\mathrm{d}$ range.
The reason for the small dispersion is the small H{\sc i} optical depth less than $\sim 0.2$ at the highest temperature, where the optically thin approximation is valid for the H{\sc i} emission.
A similar trend is also found for the local ISM in most of the high $b$ region within 200 pc of the sun \citep{2014ApJ...796...59F, 2015ApJ...798....6F}.
This suggests that the local H{\sc i} gas with the highest $T_\mathrm{d}$ shows the best correlation and is thus suited for accurate foreground subtraction in $W_\mathrm{HI}$. 

\begin{figure}
 \begin{center}
 \includegraphics[scale=1.0]{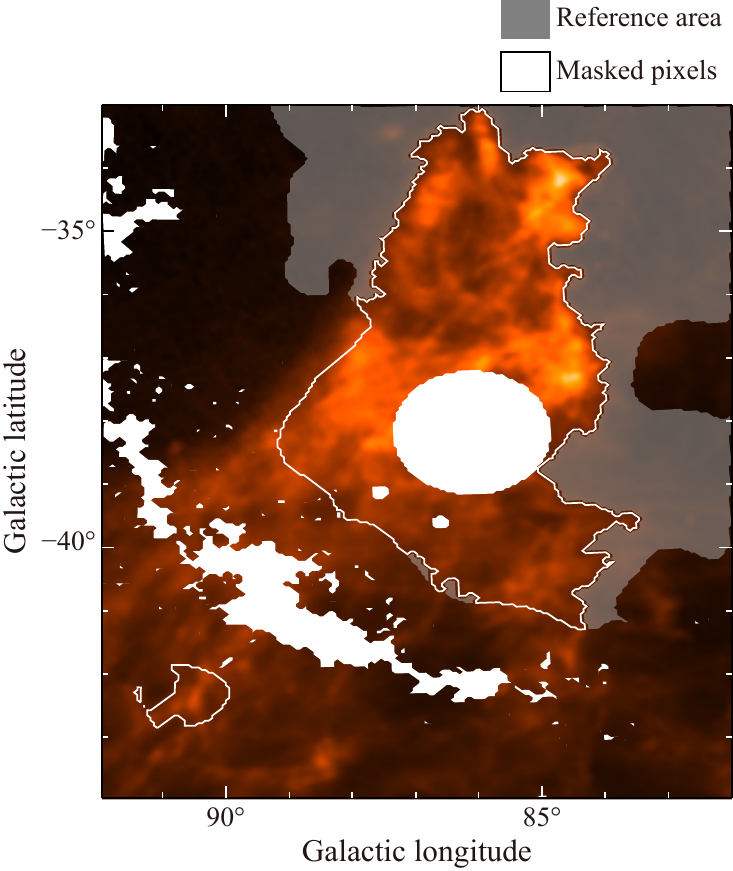}
 \end{center}
\caption{
The area of IVC 86$-$36 head (outlined by the white line) and the reference area (the gray shaded are) used in the following analyses.
White shows the masked pixels where CO emission is higher than 1.1 K km s$^{−1}$ or high $T_\mathrm{d}$ indicating localized heating by point sources (see section 2.5 of \cite{2014ApJ...796...59F}).
}\label{fig:regions}
\end{figure}

\begin{figure*}
 \begin{center}
 \includegraphics[scale=1.0]{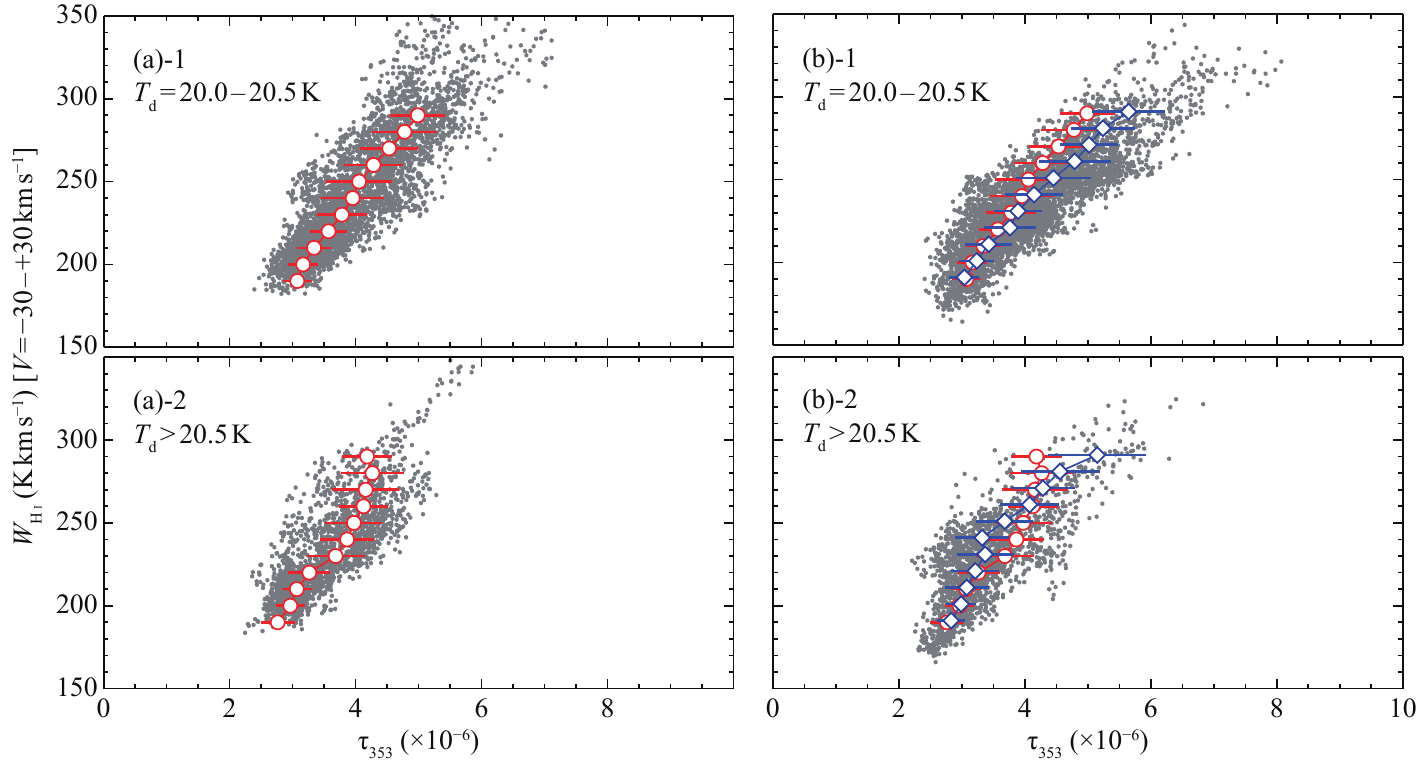}
 \end{center}
\caption{
(a) Scatter plots between $W_\mathrm{HI}$ integrated in a velocity range from $-30$ to +30 km s$^{-1}$ versus $\tau_{353}$ in the reference area.
Each panel shows for $T_\mathrm{d}$ ranges of 20.0 -- 20.5 K ((a)-1), and $T_\mathrm{d} > 20.5$ K ((a)-2).
The red circles and horizontal bars are the average and standard deviation of $\tau_{353}$ in each 10 K km s$^{-1}$ bin.
(b) Same as (a) but for the area of IVC 86$-$36 head.
The blue diamonds and bars are the average and standard deviation of $\tau_{353}$.
The red circles and bars identical to those in the Panels (a)-1–2 are superimposed for comparison.
}\label{fig:353-WHI_local_}
\end{figure*}

Figure \ref{fig:regions} shows a set up for the foreground subtraction toward IVC 86$-$36.
We chose two areas for the present analysis.
One is the area of the IVC head and the other the reference area surrounding the head with very weak H{\sc i} emission of IVC 86$-$36.
We masked regions with $T_\mathrm{d}$  lower than 20 K where H{\sc i} optical depth becomes larger than 0.2, and those with strong point sources where $\tau_{353}$ is contaminated by the infrared radiation of the point sources.
We note that the dust temperature measured by the \textit{Planck}/\textit{IRAS} shows significant increase toward IVC 86$-$36 due to the extragalactic source 3C 454.3 \citep{2009ApJ...699..817A}.
If the foreground H{\sc i} emission is uniform over the area in Figure \ref{fig:regions}, we are able to accurately subtract the foreground $\tau_{353}$ calculated from $W_\mathrm{HI}$ of the local ISM.
Figure \ref{fig:353-WHI_local_} shows scatter plots between $\tau_{353}$ and $W_\mathrm{HI}$ for $T_\mathrm{d}=20.0$ -- 20.5 K and for $T_\mathrm{d}$ higher than 20.5 K in the reference area and the area of the head.
In all the panels of Figure \ref{fig:353-WHI_local_} we see positive correlation with dispersion of $\sim 25$ \% or less in $\tau_{353}$.
The foreground ISM is therefore dominant in $\tau_{353}$ and is fairly uniformly distributed over the region shown in Figure \ref{fig:regions}.
We expect that $\tau_{353}$ of IVC 86$-$36 is obtained by subtracting the reference $\tau_{353}$ from the total $\tau_{353}$.

\begin{figure}
 \begin{center}
 \includegraphics[scale=1.0]{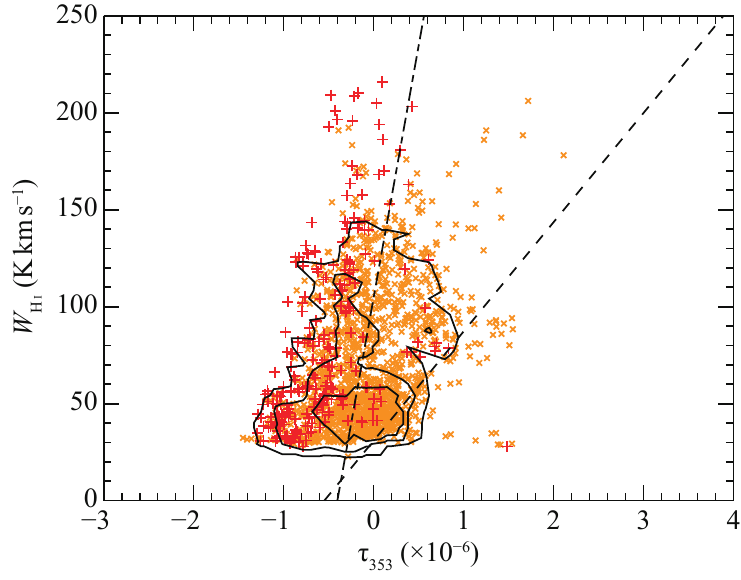}
 \end{center}
\caption{
Scatter plot between $\tau_{353}$ and $W_\mathrm{HI}$ for the area of IVC 86$-$36 head.
$\tau_{353}$ are corrected following a procedure described in Section 4.2 and $W_\mathrm{HI}$ are integrated in a velocity range from $-60$ to $-30$ km s$^{-1}$.
The orange x-marks are data points with $T_\mathrm{d}=20.5$ -- 21.0K and the red crosses are $T_\mathrm{d} > 20.5$K.
The contours include 30\%, 60\%, and 90\% of data points and the dash-dotted line is linear regressions.
The regression line for the local ISM (Figure \ref{fig:353-WHI_localISM}) is shown by the dashed line.
}\label{fig:353-WHI_IVC}
\end{figure}

The result of the subtraction is shown as a $\tau_{353}$-$W_\mathrm{HI}$ plot for the subtracted $\tau_{353}$ and $W_\mathrm{HI}$ of the IVC in Figure \ref{fig:353-WHI_IVC}.
This indicates that the IVC shows significantly smaller $\tau_{353}$ than the local ISM.
Linear regressions are obtained as follows for $T_\mathrm{d} >20.5$ K;
\begin{equation}
\textrm{the IVC:} W_\mathrm{HI} \mbox{(K km s$^{-1}$)}=(2.6\pm 0.2)\times 10^{8} \tau_{353}+(105\pm 4),
\end{equation}
and
\begin{equation}
\textrm{the local ISM:} W_\mathrm{HI} \mbox{(K km s$^{-1}$)}=(5.6\pm 0.2)\times 10^{7} \tau_{353}+(31\pm 7), 
\end{equation}
where the errors are the standard deviation obtained through the linear regression.
By taking a ratio of the two slopes we obtain $0.22\pm 0.02$ as an upper limit for the difference of the IVC relative to the local ISM.
If the dust optical depth is assumed to be proportional to the metallicity, this indicates that IVC 86$-$36 has a significantly smaller metallicity than the solar value in the local gas. 

\section{Discussion: Falling-cloud scenario}

We present a possible scenario of IVC 86$-$36.
The head-tail distribution suggests that the IVC is moving toward the Galactic plane.
It is a natural interpretation that the tail and streams were formed via dynamical interaction between the head and the halo material as shown by hydrodynamical numerical simulations for the Smith cloud \citep{2016ApJ...816L..18G}, and the tail is located behind the head which is moving toward us.
The falling velocity of the IVC is at least $\sim 60$ km s$^{-1}$ by assuming the cloud motion with an angle of 45 degrees to the line of sight.
The distance of IVC 86$-$36 is weakly constrained to be from 3 kpc to 200 pc by the stellar absorption lines and the tentative dynamical interaction signatures if real (Section 3).
If we assume a $z$-distance of 1 kpc typical to IVCs, the dynamical timescale of IVC 86$-$36 is estimated to be $\sim 1$ kpc/60 km s$^{-1}$ $\sim 20$ Myrs.

The Smith cloud shares these properties with IVC 86$-$36, which are interpreted as dynamical interaction signatures \citep{2003ApJ...591L..33L}.
The distance of the Smith cloud is 12 kpc, more distant than IVC 86$-$36, and its total mass is in excess of $10^{6}$ $M_{\solar}$ significantly larger than IVC 86$-$36.
The dynamical impact of IVC 86$-$36, e.g., the angular momentum increase as well as mass input to the Galaxy, is therefore less dramatic than the Smith cloud.
The heavy element abundance of the Smith cloud is not so anomalous as compared with the Galactic disk; absorption measurements of a sulfur spectrum argues for a Galactic disk origin of the Smith cloud, such as gas on the returning point of a fountain \citep{1980ApJ...236..577B} launched into the halo at some location in the disk \citep{2004PASJ...56..633S}.
Absorption line measurements toward the bright star HD215733 in IVC 86$-$36 at ($l$, $b$) = ($\timeform{85D}$, $\timeform{-36D}$) (Figure \ref{fig:largescaleHI}) were made at ultraviolet wavelengths by \citet{1997ApJ...475..623F}.
These authors found a trend that heavy atoms toward IVC 86$-$36 show subsolar abundance while it may be affected by the atomic ionization states and interstellar depletion.
It was thus not clear if IVC 86$-$36 originated outside the high-metallicity Galactic disk in the previous works.

The present analysis revealed that the dust optical depth of IVC 86$-$36 is less than $\sim 20$ \% as compared with that in the solar vicinity, indicating that the metal abundances of IVC 86$-$36 are less than $\sim 0.2$ solar.
Dust optical depth at sub-mm wavelength provides a measure of heavy elements that are not affected by the interstellar atomic depletion or the ionization states as long as the dust destruction process is not effective.
The interstellar shocks may affect dust destruction while the time scale is as large as 100 Myrs for large grains (e.g., \cite{1977ApJ...211L..83B, 1979ApJ...233L..81D, 1980ApJ...239..193D, 1979ApJ...231...77D, 1979ApJ...231..438D}).
It is therefore unlikely that dust destruction is important in IVC 86$-$36.

The origin of IVC 86$-$36 is of considerable interest and the metal abundances are one of the clues to pursue it.
The Magellanic Stream as well as many HVCs have metallicity in the order of $\sim 0.1$ solar according to optical observations toward bright background sources (see for a review \cite{2016ARA&A..54..363D} and references therein).
The present work showed that the metal abundances of IVC 86$-$36 are less than 0.2 solar and suggest that IVC 86$-$36 is not explained by the Galactic fountain model.
IVC 86$-$36 is located in the southern sky, and the Magellanic system may be a reasonable origin of IVC 86$-$36.
The Magellanic Stream or the SMC shows similar low metal abundance of 0.1--0.2 solar \citep{2016ARA&A..54..363D}.
The previous measurements of metal abundances were limited almost solely to the optical absorption toward compact sources in the preceding works, while the method samples a very small volume of the gas.
A large number of the background sources will help improve this deficiency.
The far infrared/sub-mm dust emission may expand the sampled volume as shown in the present work.
\citet{1987MNRAS.224.1059F} used \textit{IRAS} 100 micron emission toward the Magellanic Stream in order to quantify dust abundance in the Magellanic Stream, whereas the low sensitivity did not allow the authors to derive reliable dust abundance (see their Figure 4).
Recently, the \textit{Planck}/\textit{IRAS} data at sub-mm to 100 micron were used to derive dust abundance in the LMC and a significant variation of dust abundance was revealed in the H{\sc i} ridge of the LMC, lending support for mixing of gas with different metal abundances between the LMC and SMC, which was driven by tidal interaction between the two galaxies \citep{2017PASJ...69L...5F}.
The present work applied the same method with these authors and showed that the \textit{Planck}/\textit{IRAS} data have a potential to study dust abundance in HVCs and IVCs.

The possible physical process of the interaction of IVC 86$-$36 is a subject of considerable interest.
The formation of the long tail is probably due to the past interaction between the IVC and the halo, and the interaction is not directly detected at present.
As an example of the interaction, we find a ripple in the tail in a range from $-55$ km s$^{-1}$ to $-40$ km s$^{-1}$ at $l = \timeform{110D}$--$\timeform{90D}$ (Figure \ref{fig:momentmap}), which may suggest some instability in the frictional interaction.
The falling velocity may create shock fronts also, while we do not see signs of X-rays or thermal emission from shocked gas possibly due to cooling \citep{1998AJ....115.1693C}.
There is no detectable CO toward IVC 86$-$36 (Figure \ref{fig:dust_HI_maps}).
This requires physical modeling of the interaction while it is beyond the scope of the present paper.
It is highly desirable to make hydrodynamical numerical simulations of IVC 86$-$36 in the Galactic disk.
Questions to be answered by hydrodynamical simulations include (1) the dynamical interaction with the halo medium including heating, (2) constraints on the cloud orbit, and (3) cloud confinement and the possible role of the magnetic field, and (4) assessment of possible merging of the falling clouds into the Galactic disk. 

\section{CONCLUSIONS}

We have analyzed the Arecibo H{\sc i} data and found that an intermediate velocity cloud IVC 86$-$36 is falling onto the Galactic plane within 3 kpc of the sun.
The main conclusions are summarized below.

IVC 86$-$36 shows an intensity distribution consisting of a head-tail structure, and the tail is elongated over 40 degrees in the sky at an apparent angle of $\sim \timeform{45D}$ to the Galactic plane.
It is likely that IVC 86$-$36 is falling to the Galactic plane in a timescale of $\sim 20$ Myrs as estimated from a ratio of the length of the tail and the velocity, 1 kpc/60 km s$^{-1}$, for an assumed distance of 1 kpc.
Several parallel filamentary features streaming along the head-tail direction comprise the tail and are probably formed by the dynamical interaction between the IVC and the halo material.
Possible signatures of such interaction are shown by the bridge features etc. which are connecting IVC 86$-$36 and the disk.
The distance of IVC 86$-$36 is not well constrained, while stellar absorption places IVC 86$-$36 at a distance closer than 1--3 kpc.
The distance estimates are not in conflict with the GAIA DR2 data or soft X-ray shadows due to absorption.
The H{\sc i} mass of IVC 86$-$36 is estimated to be $7300\times (d/1\mathrm{kpc})^2$ $M_{\solar}$ under the optically thin approximation of the H{\sc i} emission.
We analyzed the H{\sc i} and \textit{Planck}/\textit{IRAS} data and found that the dust optical depth $\tau_{353}$ is too small to be detectable toward IVC 86$-$36.
For the local ISM we derived a linear relationship between $W_\mathrm{HI}$ and $\tau_{353}$ with an accuracy better than 10 \%, and we used the relationship in order to make a subtraction in $\tau_{353}$ of the foreground component, the local ISM, toward IVC 86$-$36 and derived an upper limit of the dust to H{\sc i} ratio to be less than $\sim 0.2$ solar.
This indicates that IVC 86$-$36 originated in a low metallicity environment like the Magellanic system or the halo, but not in the disk via the fountain model where the high metalicity ISM circulates in the disk.
The morphological and kinematical properties of IVC 86$-$36 are similar to the Smith cloud which is located 12 kpc away from the solar system.
An important difference from the Smith cloud is the low metallicity of IVC 86$-$36 which suggests origin of the IVC belongs to the low metallicity environment.
Detailed hydrodynamical numerical simulations of the interaction are crucial to shed more light on the physical processes of such falling gas.

\begin{ack}
This work was supported by Grants-in-Aid for Scientific Research (KAKENHI) of the Japan society for the Promotion of Science (JSPS) Grant Numbers 15H05694, 25287035. This publication utilizes data from Galactic ALFA H{\sc i} (GALFA H{\sc i}) survey data set obtained with the Arecibo L-band Feed Array (ALFA) on the Arecibo 305 m telescope. Arecibo Observatory is part of the National Astronomy and Ionosphere Center, which is operated by Cornell University under Cooperative Agreement with the U.S. National Science Foundation. The GALFA  surveys are funded by the NSF through grants to Columbia University, the University of Wisconsin, and the University of California. Some of the results in this paper have been derived using the HEALPix \citep{2005ApJ...622..759G} package.
The useful comments by the referee helped to improve the content and readability of the paper.
\end{ack}

\appendix

\section{Distance estimate by GAIA DR2}

\begin{figure}
 \begin{center}
 \includegraphics[scale=1.0]{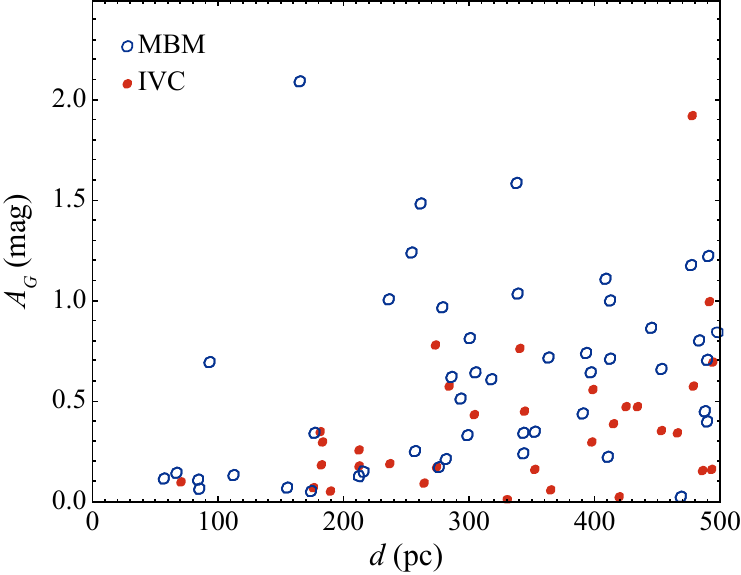}
 \end{center}
\caption{
The extinction $A_{G}$ against the distance for the GAIA sources sampled from the fields in the IVC (filled circles) and in MBM 53 (open circles).
The positions of the fields are listed in Appendix 1. 
}\label{fig:GAIA_d_AG}
\end{figure}

GAIA is an all-sky survey mission for astrometry, and the data are useful to constrain the distance of interstellar clouds. 
In addition to the parallax measurements, GAIA DR2 \citep{2018A&A...616A...1G} provides broadband photometric data of about one billion stars, giving $G$ magnitude integrated over the passband of $\sim 330$ -- 1050 nm, as well as the  low-resolution prism spectra of BP and RP \citep{2018A&A...616A...4E}.
For stars brighter than $G = 17$ mag, stellar effective temperatures, $T_\mathrm{eff}$, are obtained from the photometric data by adopting stellar color models \citep{2018A&A...616A...8A}. 
It also gives line-of-sight extinctions $A_{G}$ by using the parallaxial distances, and the reddening $E(\mathrm{BP}-\mathrm{RP})$.
We try to estimate the distance to the IVC cloud from the $A_{G}$ and distances of stars in the directions of high H{\sc i} intensity.
We chose 3 test fields with $\timeform{10'}$ radius centered at ($l$, $b$)= ($\timeform{84.7024D}$, $\timeform{-34.1755D}$), ($\timeform{85.9251D}$, $\timeform{-37.0801D}$), and ($\timeform{84.5516D}$, $\timeform{-37.3366D}$) toward peak positions of H{\sc i} intensity where the H{\sc i} is dominated by the IVC component with radial velocity of $-60$ -- $-30$ km s$^{-1}$. 
For comparison, we also chose 4 fields at ($l$, $b$) = ($\timeform{89.9009D}$, $\timeform{-33.4960D}$), ($\timeform{93.9883D}$, $\timeform{-33.1297D}$), ($\timeform{94.6707D}$, $\timeform{-35.0790D}$), and ($\timeform{95.4079D}$, $\timeform{-35.6351D}$) in MBM 53 where the H{\sc i} is dominated by the local cloud at $\sim 0$ km s$^{-1}$.
The total numbers of GAIA sources selected from the 3 fields toward the IVC and the 4 fields toward MBM 53 are 1571 and 2024, respectively. 
Figure \ref{fig:GAIA_d_AG} plots the extinction $A_{G}$ against the distance for the GAIA sources sampled from the fields in the IVC (filled circles) and in MBM 53 (open circles). 
$A_{G}$ is expected to increase with distance if the interstellar dust distributes uniformly, while it jumps if a dense interstellar cloud is located at a particular distance.
For the stars in MBM 53, a noticeable jump of $A_{G}$ by $\sim 1$ mag is seen around $d \sim 250$ pc, which nicely corresponds to the distance of MBM 53 derived by \citet{2014ApJ...786...29S}.
For those in the IVC, on the other hand, $A_{G}$ slowly increases with distance, but no clear jump is observed.
These imply that the IVC is either located farther away than 500 pc, or consist of ISM with less dust grains (i.e, metal poor condition) resulting in lower extinction. 

\section{X-ray shadows from the ROSAT data}

\begin{figure*}
 \begin{center}
 \includegraphics[scale=1.0]{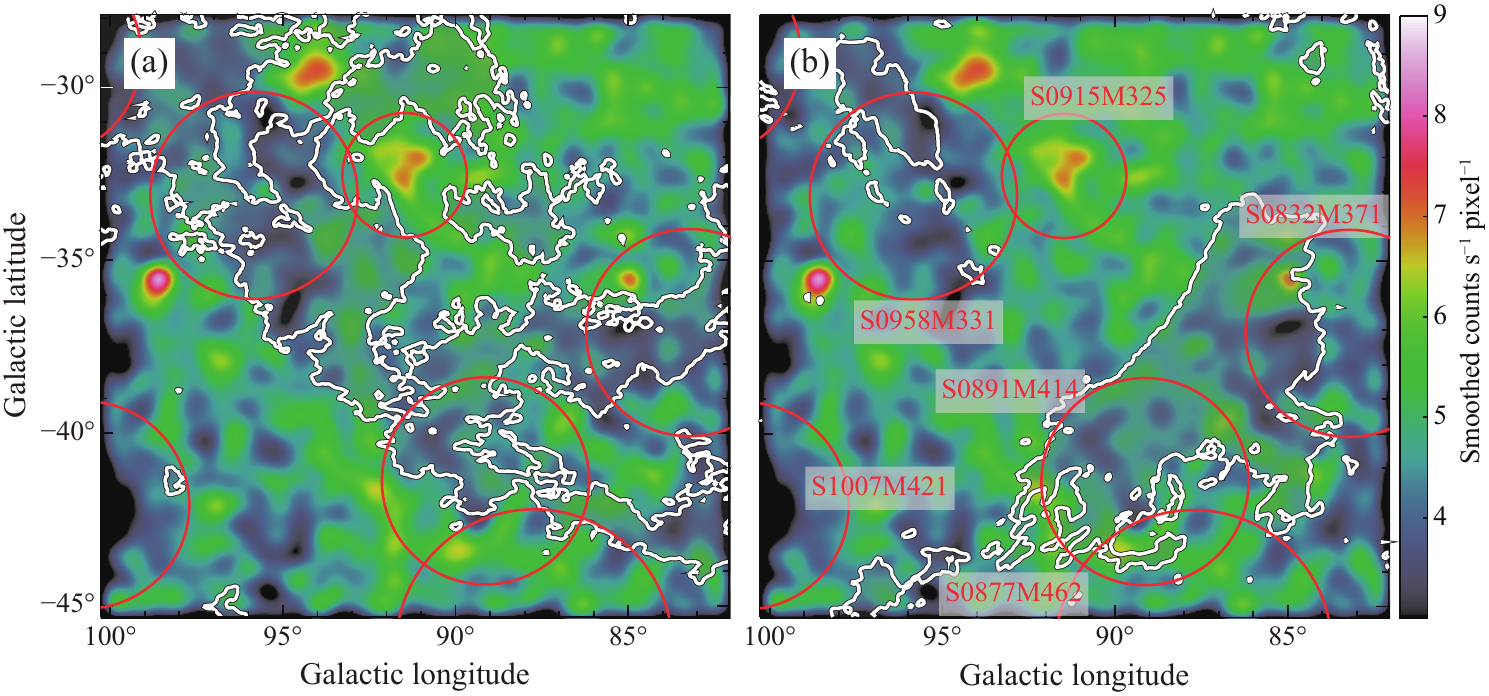}
 \end{center}
\caption{
ROSAT R1 band image toward MBM53--55/IVC 86$-$36 region.
The image is regrided to same projection type and pixel size ($\timeform{2.4'}$) as dust and H{\sc i} maps in Figure \ref{fig:dust_HI_maps}, and smoothed with a Gaussian function. 
The circles show position and size of X-ray shadows in the catalog of \citet{2000ApJS..128..171S}.
The gray-shades outlined by contours in panel (a) show the distribution of the local ISM and those in panel (b) show the IVC.
}\label{fig:Xrayshadow}
\end{figure*}

The H{\sc i} distributions of the local H{\sc i} gas and IVC 86$-$36 were compared with the soft X-ray (1/4 keV) distribution observed with ROSAT \citep{1997ApJ...485..125S}.
Figure \ref{fig:Xrayshadow}(a) shows an overlay with the local gas and Figure \ref{fig:Xrayshadow}(b) with the IVC.
The circles are taken from a list of soft X-ray shadows identified by \citet{2000ApJS..128..171S} and the names are taken from the list.
The local gas shows rough correspondence with some of the shadows, whereas the IVC shows marginal correspondence with the shadows.
It is possible that the local gas is responsible for the shadowing, although the contribution of the IVC to the shadows is uncertain.

\bibliography{reference}

\begin{thebibliography}{62}
\providecommand{\natexlab}[1]{#1}

\bibitem[{{Abdo} et~al.(2009)}]{2009ApJ...699..817A}
{Abdo}, A.~A., et~al., 2009, \emph{\apj}, 699, 817

\bibitem[{{Andrae} et~al.(2018)}]{2018A&A...616A...8A}
{Andrae}, R., et~al., 2018, \emph{\aap}, 616, A8

\bibitem[{{Barlow} \& {Silk}(1977)}]{1977ApJ...211L..83B}
{Barlow}, M.~J., \& {Silk}, J., 1977, \emph{\apjl}, 211, L83

\bibitem[{{Bland-Hawthorn} et~al.(1998){Bland-Hawthorn}, {Veilleux}, {Cecil},
  {Putman}, {Gibson}, \& {Maloney}}]{1998MNRAS.299..611B}
{Bland-Hawthorn}, J., {Veilleux}, S., {Cecil}, G.~N., {Putman}, M.~E.,
  {Gibson}, B.~K., \& {Maloney}, P.~R., 1998, \emph{\mnras}, 299, 611

\bibitem[{{Blitz} et~al.(1999){Blitz}, {Spergel}, {Teuben}, {Hartmann}, \&
  {Burton}}]{1999ApJ...514..818B}
{Blitz}, L., {Spergel}, D.~N., {Teuben}, P.~J., {Hartmann}, D., \& {Burton},
  W.~B., 1999, \emph{\apj}, 514, 818

\bibitem[{{Braun} \& {Burton}(1999)}]{1999A&A...341..437B}
{Braun}, R., \& {Burton}, W.~B., 1999, \emph{\aap}, 341, 437

\bibitem[{{Bregman}(1980)}]{1980ApJ...236..577B}
{Bregman}, J.~N., 1980, \emph{\apj}, 236, 577

\bibitem[{{Br{\"u}ns} et~al.(2000){Br{\"u}ns}, {Kerp}, {Kalberla}, \&
  {Mebold}}]{2000A&A...357..120B}
{Br{\"u}ns}, C., {Kerp}, J., {Kalberla}, P.~M.~W., \& {Mebold}, U., 2000,
  \emph{\aap}, 357, 120

\bibitem[{{Br{\"u}ns} \& {Mebold}(2004)}]{2004ASSL..312..251B}
{Br{\"u}ns}, C., \& {Mebold}, U., 2004, in H.~{van Woerden}, B.~P. {Wakker},
  U.~J. {Schwarz}, \& K.~S. {de Boer}, eds., High Velocity Clouds,
  \emph{Astrophysics and Space Science Library}, volume 312, 251

\bibitem[{{Burrows} \& {Mendenhall}(1991)}]{1991Natur.351..629B}
{Burrows}, D.~N., \& {Mendenhall}, J.~A., 1991, \emph{\nat}, 351, 629

\bibitem[{{Centurion} et~al.(1994){Centurion}, {Vladilo}, {de Boer},
  {Herbstmeier}, \& {Schwarz}}]{1994A&A...292..261C}
{Centurion}, M., {Vladilo}, G., {de Boer}, K.~S., {Herbstmeier}, U., \&
  {Schwarz}, U.~J., 1994, \emph{\aap}, 292, 261

\bibitem[{{Clarke} et~al.(2001){Clarke}, {Gladders}, \&
  {Mall{\'e}n-Ornelas}}]{2001ASPC..240..529C}
{Clarke}, T.~E., {Gladders}, M., \& {Mall{\'e}n-Ornelas}, G., 2001, in J.~E.
  {Hibbard}, M.~{Rupen}, \& J.~H. {van Gorkom}, eds., Gas and Galaxy Evolution,
  \emph{Astronomical Society of the Pacific Conference Series}, volume 240, 529

\bibitem[{{Condon} et~al.(1998){Condon}, {Cotton}, {Greisen}, {Yin}, {Perley},
  {Taylor}, \& {Broderick}}]{1998AJ....115.1693C}
{Condon}, J.~J., {Cotton}, W.~D., {Greisen}, E.~W., {Yin}, Q.~F., {Perley},
  R.~A., {Taylor}, G.~B., \& {Broderick}, J.~J., 1998, \emph{\aj}, 115, 1693

\bibitem[{{Connors} et~al.(2006){Connors}, {Kawata}, {Bailin}, {Tumlinson}, \&
  {Gibson}}]{2006ApJ...646L..53C}
{Connors}, T.~W., {Kawata}, D., {Bailin}, J., {Tumlinson}, J., \& {Gibson},
  B.~K., 2006, \emph{\apjl}, 646, L53

\bibitem[{{D'Onghia} \& {Fox}(2016)}]{2016ARA&A..54..363D}
{D'Onghia}, E., \& {Fox}, A.~J., 2016, \emph{\araa}, 54, 363

\bibitem[{{Draine}(2011)}]{2011piim.book.....D}
{Draine}, B.~T., 2011, {Physics of the Interstellar and Intergalactic Medium}
  (Princeton, NJ: Princeton University Press)

\bibitem[{{Draine} \& {Salpeter}(1979{\natexlab{a}})}]{1979ApJ...231..438D}
{Draine}, B.~T., \& {Salpeter}, E.~E., 1979{\natexlab{a}}, \emph{\apj}, 231,
  438

\bibitem[{{Draine} \& {Salpeter}(1979{\natexlab{b}})}]{1979ApJ...231...77D}
{Draine}, B.~T., \& {Salpeter}, E.~E., 1979{\natexlab{b}}, \emph{\apj}, 231, 77

\bibitem[{{Dwek} \& {Scalo}(1979)}]{1979ApJ...233L..81D}
{Dwek}, E., \& {Scalo}, J.~M., 1979, \emph{\apjl}, 233, L81

\bibitem[{{Dwek} \& {Scalo}(1980)}]{1980ApJ...239..193D}
{Dwek}, E., \& {Scalo}, J.~M., 1980, \emph{\apj}, 239, 193

\bibitem[{{Evans} et~al.(2018)}]{2018A&A...616A...4E}
{Evans}, D.~W., et~al., 2018, \emph{\aap}, 616, A4

\bibitem[{{Fitzgerald}(1970)}]{1970A&A.....4..234F}
{Fitzgerald}, M.~P., 1970, \emph{\aap}, 4, 234

\bibitem[{{Fitzpatrick} \& {Spitzer}(1997)}]{1997ApJ...475..623F}
{Fitzpatrick}, E.~L., \& {Spitzer}, Jr., L., 1997, \emph{\apj}, 475, 623

\bibitem[{{Fong} et~al.(1987){Fong}, {Jones}, {Shanks}, {Stevenson}, \&
  {Strong}}]{1987MNRAS.224.1059F}
{Fong}, R., {Jones}, L.~R., {Shanks}, T., {Stevenson}, P.~R.~F., \& {Strong},
  A.~W., 1987, \emph{\mnras}, 224, 1059

\bibitem[{{Fox} et~al.(2016)}]{2016ApJ...816L..11F}
{Fox}, A.~J., et~al., 2016, \emph{\apjl}, 816, L11

\bibitem[{{Fukui} et~al.(2018{\natexlab{a}}){Fukui}, {Hayakawa}, {Inoue},
  {Torii}, {Okamoto}, {Tachihara}, {Onishi}, \&
  {Hayashi}}]{2018ApJ...860...33F}
{Fukui}, Y., {Hayakawa}, T., {Inoue}, T., {Torii}, K., {Okamoto}, R.,
  {Tachihara}, K., {Onishi}, T., \& {Hayashi}, K., 2018{\natexlab{a}},
  \emph{\apj}, 860, 33

\bibitem[{{Fukui} et~al.(2015){Fukui}, {Torii}, {Onishi}, {Yamamoto},
  {Okamoto}, {Hayakawa}, {Tachihara}, \& {Sano}}]{2015ApJ...798....6F}
{Fukui}, Y., {Torii}, K., {Onishi}, T., {Yamamoto}, H., {Okamoto}, R.,
  {Hayakawa}, T., {Tachihara}, K., \& {Sano}, H., 2015, \emph{\apj}, 798, 6

\bibitem[{{Fukui} et~al.(2017){Fukui}, {Tsuge}, {Sano}, {Bekki}, {Yozin},
  {Tachihara}, \& {Inoue}}]{2017PASJ...69L...5F}
{Fukui}, Y., {Tsuge}, K., {Sano}, H., {Bekki}, K., {Yozin}, C., {Tachihara},
  K., \& {Inoue}, T., 2017, \emph{\pasj}, 69, L5

\bibitem[{{Fukui} et~al.(2014)}]{2014ApJ...796...59F}
{Fukui}, Y., et~al., 2014, \emph{\apj}, 796, 59

\bibitem[{{Fukui} et~al.(2018{\natexlab{b}})}]{2018ApJ...859..166F}
{Fukui}, Y., et~al., 2018{\natexlab{b}}, \emph{\apj}, 859, 166

\bibitem[{{Gaia Collaboration}(2018)}]{2018A&A...616A...1G}
{Gaia Collaboration}, 2018, \emph{\aap}, 616, A1

\bibitem[{{Galyardt} \& {Shelton}(2016)}]{2016ApJ...816L..18G}
{Galyardt}, J., \& {Shelton}, R.~L., 2016, \emph{\apjl}, 816, L18

\bibitem[{{G{\'o}rski} et~al.(2005){G{\'o}rski}, {Hivon}, {Banday}, {Wandelt},
  {Hansen}, {Reinecke}, \& {Bartelmann}}]{2005ApJ...622..759G}
{G{\'o}rski}, K.~M., {Hivon}, E., {Banday}, A.~J., {Wandelt}, B.~D., {Hansen},
  F.~K., {Reinecke}, M., \& {Bartelmann}, M., 2005, \emph{\apj}, 622, 759

\bibitem[{{Lesh}(1968)}]{1968ApJS...17..371L}
{Lesh}, J.~R., 1968, \emph{\apjs}, 17, 371

\bibitem[{{Lockman}(1984)}]{1984ApJ...283...90L}
{Lockman}, F.~J., 1984, \emph{\apj}, 283, 90

\bibitem[{{Lockman}(2003)}]{2003ApJ...591L..33L}
{Lockman}, F.~J., 2003, \emph{\apjl}, 591, L33

\bibitem[{{Lockman} et~al.(2007){Lockman}, {Benjamin}, {Heroux}, \&
  {Langston}}]{2007AAS...21114801L}
{Lockman}, F.~J., {Benjamin}, R.~A., {Heroux}, A., \& {Langston}, G.~I., 2007,
  in American Astronomical Society Meeting Abstracts, \emph{Bulletin of the
  American Astronomical Society}, volume~39, 1000

\bibitem[{{Lockman} et~al.(2002){Lockman}, {Murphy}, {Petty-Powell}, \&
  {Urick}}]{2002ApJS..140..331L}
{Lockman}, F.~J., {Murphy}, E.~M., {Petty-Powell}, S., \& {Urick}, V.~J., 2002,
  \emph{\apjs}, 140, 331

\bibitem[{{Magnani} et~al.(1985){Magnani}, {Blitz}, \&
  {Mundy}}]{1985ApJ...295..402M}
{Magnani}, L., {Blitz}, L., \& {Mundy}, L., 1985, \emph{\apj}, 295, 402

\bibitem[{{Maller} \& {Bullock}(2004)}]{2004MNRAS.355..694M}
{Maller}, A.~H., \& {Bullock}, J.~S., 2004, \emph{\mnras}, 355, 694

\bibitem[{{Mermilliod}(1987)}]{1987A&AS...71..119M}
{Mermilliod}, J.-C., 1987, \emph{\aaps}, 71, 119

\bibitem[{{Mirabel} \& {Morras}(1990)}]{1990ApJ...356..130M}
{Mirabel}, I.~F., \& {Morras}, R., 1990, \emph{\apj}, 356, 130

\bibitem[{{Miville-Desch{\^e}nes} \& {Lagache}(2005)}]{2005ApJS..157..302M}
{Miville-Desch{\^e}nes}, M.-A., \& {Lagache}, G., 2005, \emph{\apjs}, 157, 302

\bibitem[{{Nichols} \& {Bland-Hawthorn}(2009)}]{2009ApJ...707.1642N}
{Nichols}, M., \& {Bland-Hawthorn}, J., 2009, \emph{\apj}, 707, 1642

\bibitem[{{Peek} et~al.(2007){Peek}, {Putman}, {McKee}, {Heiles}, \&
  {Stanimirovi{\'c}}}]{2007ApJ...656..907P}
{Peek}, J.~E.~G., {Putman}, M.~E., {McKee}, C.~F., {Heiles}, C., \&
  {Stanimirovi{\'c}}, S., 2007, \emph{\apj}, 656, 907

\bibitem[{{Peek} et~al.(2011)}]{2011ApJS..194...20P}
{Peek}, J.~E.~G., et~al., 2011, \emph{\apjs}, 194, 20

\bibitem[{{Planck Collaboration}(2014{\natexlab{a}})}]{2014A&A...571A...6P}
{Planck Collaboration}, 2014{\natexlab{a}}, \emph{\aap}, 571, A6

\bibitem[{{Planck Collaboration}(2014{\natexlab{b}})}]{2014A&A...571A..11P}
{Planck Collaboration}, 2014{\natexlab{b}}, \emph{\aap}, 571, A11

\bibitem[{{Richter}(2017)}]{2017ASSL..430...15R}
{Richter}, P., 2017, in A.~{Fox} \& R.~{Dav{\'e}}, eds., Gas Accretion onto
  Galaxies, \emph{Astrophysics and Space Science Library}, volume 430, 15

\bibitem[{{Schlafly} et~al.(2014)}]{2014ApJ...786...29S}
{Schlafly}, E.~F., et~al., 2014, \emph{\apj}, 786, 29

\bibitem[{{Schlegel} et~al.(1998){Schlegel}, {Finkbeiner}, \&
  {Davis}}]{1998ApJ...500..525S}
{Schlegel}, D.~J., {Finkbeiner}, D.~P., \& {Davis}, M., 1998, \emph{\apj}, 500,
  525

\bibitem[{{Smith}(1963)}]{1963BAN....17..203S}
{Smith}, G.~P., 1963, \emph{\bain}, 17, 203

\bibitem[{{Snowden} et~al.(2000){Snowden}, {Freyberg}, {Kuntz}, \&
  {Sanders}}]{2000ApJS..128..171S}
{Snowden}, S.~L., {Freyberg}, M.~J., {Kuntz}, K.~D., \& {Sanders}, W.~T., 2000,
  \emph{\apjs}, 128, 171

\bibitem[{{Snowden} et~al.(1991){Snowden}, {Mebold}, {Hirth}, {Herbstmeier}, \&
  {Schmitt}}]{1991Sci...252.1529S}
{Snowden}, S.~L., {Mebold}, U., {Hirth}, W., {Herbstmeier}, U., \& {Schmitt},
  J.~H.~M., 1991, \emph{Science}, 252, 1529

\bibitem[{{Snowden} et~al.(1997)}]{1997ApJ...485..125S}
{Snowden}, S.~L., et~al., 1997, \emph{\apj}, 485, 125

\bibitem[{{Sofue} et~al.(2004){Sofue}, {Kudoh}, {Kawamura}, {Shibata}, \&
  {Fujimoto}}]{2004PASJ...56..633S}
{Sofue}, Y., {Kudoh}, T., {Kawamura}, A., {Shibata}, K., \& {Fujimoto}, M.,
  2004, \emph{\pasj}, 56, 633

\bibitem[{{Takahira} et~al.(2014){Takahira}, {Tasker}, \&
  {Habe}}]{2014ApJ...792...63T}
{Takahira}, K., {Tasker}, E.~J., \& {Habe}, A., 2014, \emph{\apj}, 792, 63

\bibitem[{{van Woerden} et~al.(2004){van Woerden}, {Wakker}, {Schwarz}, \& {de
  Boer}}]{2004ASSL..312.....V}
{van Woerden}, H., {Wakker}, B.~P., {Schwarz}, U.~J., \& {de Boer}, K.~S.,
  eds., 2004, {High Velocity Clouds}, \emph{Astrophysics and Space Science
  Library}, volume 312

\bibitem[{{Wakker}(2001)}]{2001ApJS..136..463W}
{Wakker}, B.~P., 2001, \emph{\apjs}, 136, 463

\bibitem[{{Wakker} \& {van Woerden}(1997)}]{1997ARA&A..35..217W}
{Wakker}, B.~P., \& {van Woerden}, H., 1997, \emph{\araa}, 35, 217

\bibitem[{{Walborn}(1976)}]{1976ApJ...205..419W}
{Walborn}, N.~R., 1976, \emph{\apj}, 205, 419

\bibitem[{{Yamamoto} et~al.(2003){Yamamoto}, {Onishi}, {Mizuno}, \&
  {Fukui}}]{2003ApJ...592..217Y}
{Yamamoto}, H., {Onishi}, T., {Mizuno}, A., \& {Fukui}, Y., 2003, \emph{\apj},
  592, 217

\end{thebibliography}

\newpage

%%%%%%%%%%%%%%%%%%%%%%%%%%%%%%%%%%%%%%%%%%%%%%%%%%%%%%%%%%%%%%%%%%%%%%%%%%%%%%%%%%%%%%%%%%

\end{document}